\documentclass[journal]{IEEEtran}

\usepackage{amsmath,bm,amssymb,amsfonts,amsthm}
\usepackage{algorithm,algorithmic}
\usepackage{graphicx}
\usepackage{xcolor}
\usepackage{relsize}
\usepackage{textcomp}
\usepackage{listings}

\usepackage{changepage}
\usepackage{hhline}
\usepackage{multirow}
\usepackage{nomencl}
\usepackage{mathtools}
\usepackage{commath}
\usepackage{booktabs}
\usepackage{subfiles}
\usepackage{tabularx}
\usepackage{lscape}
\usepackage{rotating}

\usepackage[normalem]{ulem}
\usepackage[utf8]{inputenc}
\usepackage[hyphens]{url}
\usepackage{lineno}
    \modulolinenumbers[5]
\usepackage[caption=false,font=footnotesize]{subfig}
\usepackage{float}
\usepackage{enumitem}
    \setlist{nolistsep}
\usepackage{scalerel}
\usepackage{tikz}
\usetikzlibrary{svg.path}
\definecolor{orcidlogocol}{HTML}{A6CE39}
\tikzset{
  orcidlogo/.pic={
    \fill[orcidlogocol] svg{M256,128c0,70.7-57.3,128-128,128C57.3,256,0,198.7,0,128C0,57.3,57.3,0,128,0C198.7,0,256,57.3,256,128z};
    \fill[white] svg{M86.3,186.2H70.9V79.1h15.4v48.4V186.2z}
                 svg{M108.9,79.1h41.6c39.6,0,57,28.3,57,53.6c0,27.5-21.5,53.6-56.8,53.6h-41.8V79.1z M124.3,172.4h24.5c34.9,0,42.9-26.5,42.9-39.7c0-21.5-13.7-39.7-43.7-39.7h-23.7V172.4z}
                 svg{M88.7,56.8c0,5.5-4.5,10.1-10.1,10.1c-5.6,0-10.1-4.6-10.1-10.1c0-5.6,4.5-10.1,10.1-10.1C84.2,46.7,88.7,51.3,88.7,56.8z};
  }
}
\newcommand\orcidicon[1]{\href{https://orcid.org/#1}{\mbox{\scalerel*{
\begin{tikzpicture}[yscale=-1,transform shape]
\pic{orcidlogo};
\end{tikzpicture}
}{|}}}}
\usepackage{hyperref}

\usepackage[noadjust]{cite}

\begin{document}

\title{\huge Impacts of Earthquakes on Electrical Grid Resilience}

\author{
    Adam~Mate $^{1}$\orcidicon{0000-0002-5628-6509} \IEEEmembership{Member, IEEE}
    and~Travis~Hagan $^{2}$ \IEEEmembership{Student Member, IEEE}
    and\\~Eduardo~Cotilla-Sanchez $^{2}$\orcidicon{0000-0002-3964-3260} \IEEEmembership{Senior Member, IEEE}
    and~Ted~K.~A.~Brekken $^{2}$\orcidicon{0000-0002-3093-108X} \IEEEmembership{Senior Member, IEEE}
    and\\~Annette~Von~Jouanne $^{3}$ \IEEEmembership{Senior Member, IEEE}

\thanks{$^{1}$ The author is with the Advanced Network Science Initiative at Los Alamos National Laboratory (LANL ANSI), Los Alamos, NM 87544 USA. Email:\{amate\}@lanl.gov.}

\thanks{$^{2}$ The authors are with the School of Electrical Engineering and Computer Science, Oregon State University, Corvallis, OR 97331 USA. Email:\{hagantr, ecs, brekken\}@oregonstate.edu.}

\thanks{$^{3}$ The author is with the Energy Systems Department of Electrical and Computer Engineering, Baylor University, Waco, TX 76706 USA. Email:\{annette\_vonjouanne\}@baylor.edu.}

\thanks{Color versions of one or more of the figures in this paper are available online at https://ieeexplore.ieee.org.}

}

\markboth{IEEE/IAS 57th Industrial \& Commercial Power Systems Technical Conference, April~2021}{}

\maketitle

\begin{abstract}
One of the most complex and devastating disaster scenarios that the U.S.~Pacific Northwest region and the state of Oregon faces is a large magnitude Cascadia Subduction Zone earthquake event.
The region's electrical grid lacks in resilience against the destruction of a megathrust earthquake, a powerful tsunami, hundreds of aftershocks and increased volcanic activity, all of which are highly probable components of this hazard.\\
This research seeks to catalyze further understanding and improvement of resilience.
By systematizing power system related experiences of historical earthquakes, and collecting practical and innovative ideas from other regions on how to enhance network design, construction, and operation, important steps are being taken toward a more resilient, earthquake-resistant grid.
This paper presents relevant findings in an effort to be an overview and a useful guideline for those who are also working towards greater electrical grid resilience.

\end{abstract}


\section{Introduction}

Since the 1980s, when researchers recognized the Cascadia Subduction Zone (abbr.~CSZ) as an active fault, the scientific community has been aware of the major geological hazard that the state of Oregon and the U.S.~Pacific Northwest region (abbr.~PNW) faces:~the possibility of a tremendous earthquake and tsunami caused by this fault, which can strike at any moment and can cause enormous destruction thorough the region \cite{ref-01}.

The Oregon Talent Council (abbr.~OTC), in their Oregon Talent Plan \cite{ref-02}, identified a critical need for more and better prepared electrical power systems engineers to serve Oregon's energy technologies and utility industry.
There is an urgent need for agile engineers with expertise combining technical education, applied skills, and work experience focused in disaster preparedness and system resilience.
Consequently, OTC decided to fund the \emph{Pacific Northwest Electrical System Resiliency and Disaster Preparedness Training Project} to address the existing talent gap; to conduct novel research on electrical system resilience, to increase resilience awareness, and to train Oregon electrical engineers for disaster preparedness with particular attention to a CSZ event. 

This paper presents some of the key findings of the OTC funded research.
It aims to systematize power system related experiences of historical catastrophic earthquakes in order to gain knowledge that can help the PNW to best prepare for a future CSZ event.
The key contribution is a comprehensive collection of proven practices and innovative ideas from other regions and nations of the world, on how to improve power system design, construction and operation, which can help to transform Oregon’s and the PNW’s  electrical grid into a more resilient, earthquake and tsunami resistant power system.

The remainder of this paper is organized as follows.
Section~II. gives a comprehensive review on the topic’s background:~information about the CSZ and its future, current resilience in Oregon and efforts to increase it, and a brief introduction of historical earthquakes that can serve as exemplar cases to provide valuable input.
Sections~III., IV., and V. present findings related to the generation, transmission and distribution levels of the electrical grid, respectively.
Section~VI summarizes the research and concludes with future steps and foreseeable difficulties.

\section{Background}

\subsection{Next Cascadia Earthquake}

The CSZ is an approx.~600 miles long offshore fault, which lies in the coastal region of the PNW, stretching from Cape Mendocino in northern California -- through the states of Oregon and Washington -- to the Brooks Peninsula in southern British Columbia \cite{ref-01, ref-03, ref-04, ref-05}.
The fault is part of a great arc of subduction zones that surrounds the Pacific Ocean, creating a formation called the ``Ring of Fire,'' and is a geologic mirror image of the subduction zone lying east of Japan \cite{ref-01}.

\begin{figure}
\centerline{\includegraphics[scale=0.175]{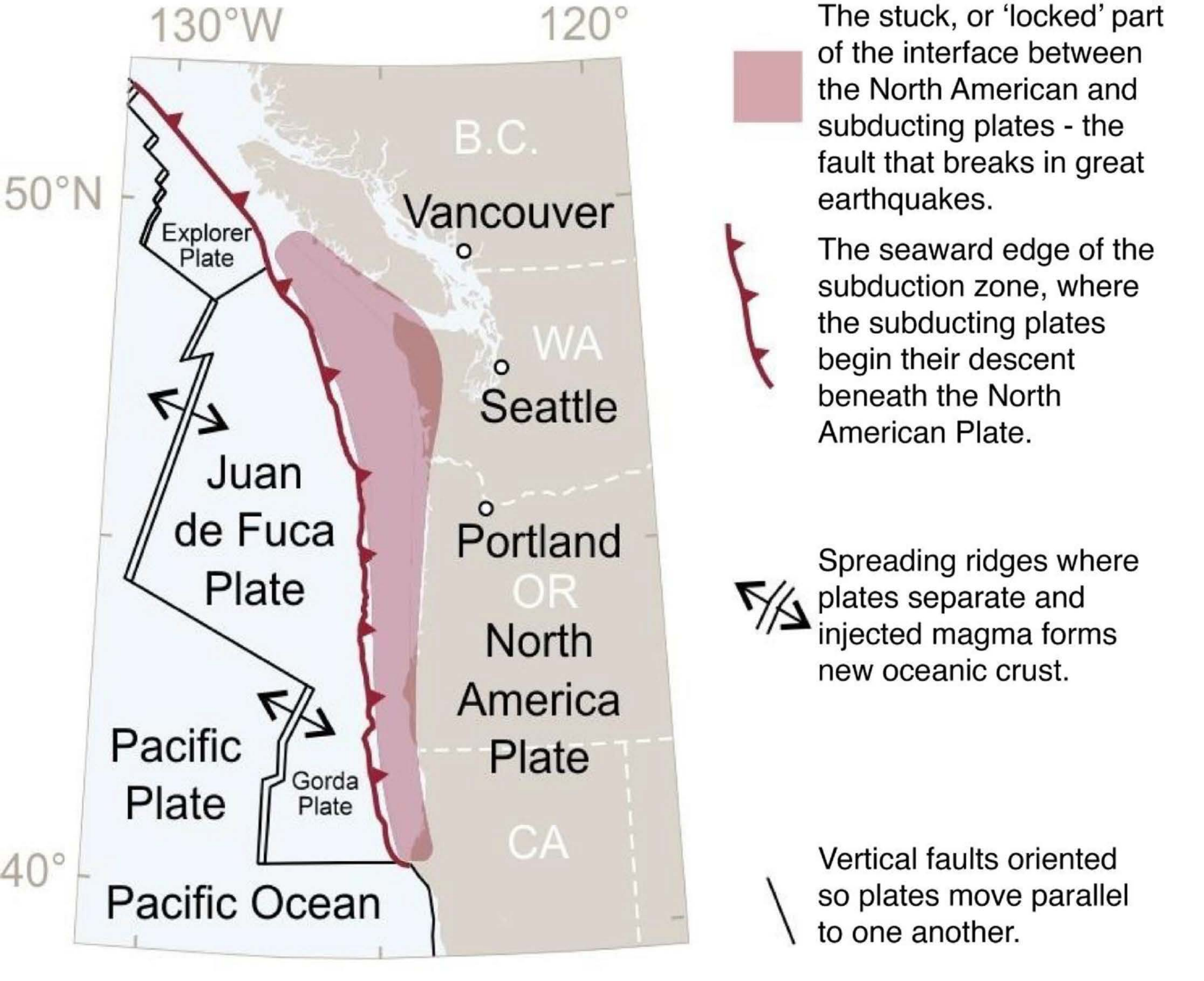}}
\caption{Schematic view of the tectonic plates in the CSZ area \cite{ref-05, ref-06}.}
\label{CSZ_map}
\end{figure}

In the CSZ, three denser oceanic tectonic plates (namely the Explorer, Juan de Fuca, and Gorda plates) are sliding from west to east and subducting beneath the less dense continental plate (North American plate) that moves in a general southwest direction, overriding the oceanic plates. Fig.~\ref{CSZ_map} illustrates the location of the subduction zone and these plates.

The movement of these plates is neither constant nor smooth:~the plates stick, building stress until the fault suddenly breaks and releases the accumulated energy in the form of an earthquake(s) \cite{ref-03}. Thereafter, the plates start moving again, and continue to move, until getting stuck again.

There is no doubt that another subduction earthquake will strike the PNW in the future. In the last decade, research has confirmed that the CSZ has a long history of great earthquakes.
The most recent happened on January 26, 1700, creating a magnitude 9.0 earthquake followed within minutes by a large tsunami.
Energy for the next earthquake is currently building up along the fault, and has been since the last earthquake \cite{ref-01, ref-03, ref-05}.
The time interval between previous CSZ events varied from a few decades to many centuries, but most intervals were shorter than the time elapsed since the last event in 1700 \cite{ref-01}. The calculated odds that the next earthquake will occur in the next 50 years range from 7-15\% for a ``great'' earthquake affecting the entire PNW to about 37\% for a ``very large'' earthquake affecting southern Oregon and northern California \cite{ref-01, ref-03, ref-04, ref-05}.

Geologists assembled a ten-thousand-year record of past events, by studying sediments in coastal marshes and on the ocean floor.
This shows that half of the past earthquakes have been ``very large'' (estimated magnitude of 8.3 to 8.6) and centered on the southern Oregon coast, while the other half have been ``great'' (estimated magnitude 8.7 to 9.3) and extended along the full length of the fault \cite{ref-01}.
Although it is possible that the next CSZ earthquake(s) will be a partial rupture of the fault, section by section, in a series of large events over a period of years, it is strongly anticipated by many scientists that it will be similar to the last event in 1700, and will be the result of the entire fault rupturing, causing one great earthquake measuring magnitude 9.0, with ground shaking lasting 4-6 minutes \cite{ref-03, ref-05}.

\subsection{Earthquake resilience in Oregon}

After the discovery of the major threat, the rate of change to increase Oregon’s resilience has been slow.
The first-time that explicit seismic provisions were adopted in Oregon’s building codes was in 1993 \cite{ref-01, ref-07}.
Only following the 2011 Tohoku, Japan earthquake and tsunami disaster, was special attention given to the CSZ threat at the state governmental level.
The Oregon Seismic Safety Policy Advisory Commission, with the help of more than 150 volunteer professionals, prepared the Oregon Resilience Plan in 2013 \cite{ref-01}. This plan was the first comprehensive study and had the following goals:~assess the state of resilience in Oregon, plan for the impacts of a CSZ event, and map a path of policy and investment priorities for the next fifty years.
Up until today, two CSZ earthquake scenario documents have been created, \cite{ref-03} and \cite{ref-05} (with multiple versions), with the goal to provide information for the public on the hazards the PNW faces.

These three studies \cite{ref-01, ref-03, ref-05} identified the following key findings on the current state of resilience of Oregon’s electrical grid:
\begin{itemize}
\item Electrical facilities and network components -- including power plants, substations, transmission lines -- are seismically vulnerable to damage and have significant risk due to ground shaking and ground failure, especially landslides, soil liquefaction, lateral spreading and coastal subsidence hazards.
\item Most of Oregon’s critical and non-critical energy infrastructure has been constructed with seismic design deficiencies, and not initially built to with-stand earthquakes. Substantial improvements, investments, and a uniform set of design and construction codes are needed to minimize extensive direct earthquake damage, indirect losses, and possible ripple effects.
\item A CSZ event will cause the failure of numerous power system components, and over half of the region’s electrical grid may suffer medium to high damage on all grid-levels. Outages and blackouts may occur not only within 100 miles of the coastline, but even in areas that were not directly affected by the earthquake. The destruction on the coastal areas may be severe enough as to render the equipment and structures irreparable. In the I-5 corridor, considerable damage on generation and distribution levels may result in the loss of over half of the system’s capacity.
\item Oregon’s liquid fuel supply is extremely vulnerable. Besides the high dependency on Washington State -- more than 90\% of Oregon’s refined petroleum products come from the Puget Sound area, which is also vulnerable to a CSZ earthquake -- the storage facilities along Oregon’s Willamette River lie on liquefiable riverside soils. This significantly affects most sectors of the economy critical to emergency response and economic recovery.
\item Estimated restoration time of electrical grid after a CSZ event ranges from 1-3 months (in the Willamette Valley) to 3-6 months (Coastal region), depending on the degree of destruction, available utility personnel, contractors and road conditions.
\end{itemize}

Even though small steps -- conducted studies, funded research, legislative changes –- have been taken in the recent years to start and accelerate the needed change in this issue, Oregon’s electrical grid today is still far from resilient to the impacts of a CSZ event.

\subsection{Notable Historical Earthquakes}

\begin{table*}
\caption{Comparison of Historical Catastrophic Earthquakes}
\label{Earthquake_table}
\includegraphics[width=\linewidth]{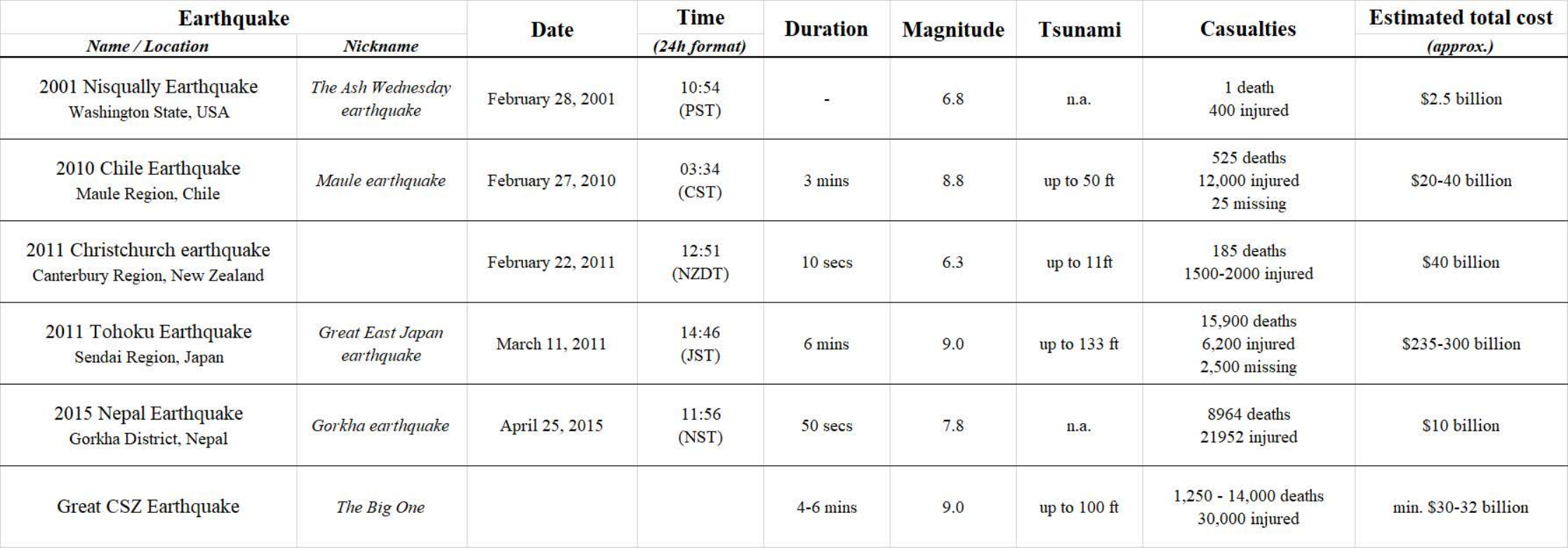}
\end{table*}

Other regions and nations of the world can provide plenty of valuable firsthand experiences, useful practices and ideas for the PNW to utilize.
Even though environmental and geological conditions and occurring natural disasters will be unique to those locations, there are methods, thoughts, tools and developments that can be applied to make the PNW’s electric grid more resilient.

Table~\ref{Earthquake_table}. presents historical catastrophic earthquakes (in chronological order) that were investigated in this research.
Besides comparing the main attributes of these events to each other, the last line also compares them to the upcoming CSZ earthquake. Thus, the magnitude of this future event becomes more easily conceivable.
The data in the table is compiled from sources listed in the References section and publicly available information.

The PNW, and specifically the state of Oregon, is prone to earthquakes, just like all other regions inside the Pacific Ring of Fire.
In this research special attention was paid to Japan, Chile and New Zealand. Earthquakes of all sizes are much more common in these areas than anywhere else in the world, and as such, these countries have become better prepared and possess years of expertise in electrical grid resilience related questions.
The Nisqually earthquake is one of the few earthquakes in the PNW that was near a major metropolitan area and a detailed analysis of the power grid had been completed.
The Nepal earthquake was included because it shows the effect of earthquakes on mountainous regions -- similar to some areas of the PNW -- and the additional considerations that must be made when dealing with this type of geography.

The following three sections present findings that are divided into the categories of generation, transmission, and distribution levels of the electrical grid.

\section{Generation Level}

Generation plants across the PNW face a varying degree of threat from landslides, soil liquefaction, ground settling, lateral spreading and ground shaking. Each threat has the potential to take generation plants off-line for months.

Landslides are more unique to the geography of the PNW. The relatively wet winters combined with the number of hills and mountains creates an environment where events of this nature are expected.
The 7.8 magnitude earthquake that struck Nepal in 2015, caused tens of thousands of landslides, and illustrates the link between earthquakes and landslides \cite{ref-08}.
Saturated soils can have about half the strength than comparable dry soils \cite{ref-08}. This is concerning for the Willamette Valley in Oregon, Puget Sound in Washington, coastal regions and the Columbia River Gorge \cite{ref-08}, which is where the majority of the generation facilities lie.

The PNW has several hydroelectric dams spanning the U.S. states of Oregon, Washington, Idaho, and in the Canadian Province of British Columbia. These dams supply about 75\% of the PNW's power demands, and some of the dams lie in areas prone to landslides.
The Columbia River Gorge also contains most of the PNW’s wind generation. The wind turbines, perched on hills above the river, are also at risk from landslides from a large magnitude earthquake.
A small slide near the Columbia River has the potential to vastly affect and take down generation capabilities.

The worst-case scenario is a slide of the same magnitude as the “Bonneville slide”.
This slide, which is believed to be the result of an earthquake between 1550 and 1750, dammed the Columbia River near present day Bonneville Dam, and created a lake spanning 100 miles upstream. Sometime later, the dam breached and created the Cascades Rapids. This feature is now hidden by Bonneville Reservoir \cite{ref-09}.
If the same event occurred today, Bonneville Dam would be completely buried, and the Dalles Dam and John Day Dam would have no outflow. This would lead to a loss of nearly five gigawatts of nameplate capacity directly, and several more gigawatts indirectly.

Soil liquefaction, settling, and lateral spreading, and the phenomena that they cause, have great potential for damage in the PNW. They are best studied and understood through analysis of earthquakes in other countries.
In the Nepal and Tohoku, Japan earthquakes, most of the settling –- induced by soil liquefaction –- occurred near the coastline and in reclaimed soil \cite{ref-10}.
As such, generation sites that lie near water sources will experience a greater amount of settling, leading to more damage at these facilities. This poses a problem for generation along the Columbia River especially \cite{ref-08}.

Ground shaking is the underlying cause behind landslides, soil liquefaction, settling, and lateral spreading, but it also poses a direct risk to generation sites.
In Japan, many generation sources went offline due to the shifting of vibration isolators \cite{ref-10}.
Similar issues were encountered in the New Zealand earthquakes where mercury safety switches disconnected transformers due to shaking \cite{ref-11}.
Ground shaking is particularly an issue in concrete structures, which can crack under the repeated stress \cite{ref-11}.

Another important consideration is the survivability of dams. Many of the dams in the northwest were built prior to understanding the CSZ, and seismic effects were not considered. 
Despite this, dams have been shown to hold up well in earthquakes \cite{ref-12}. According to a PUD (Public Utility District) in Washington State that oversees several of the dams, the dams are robust and will survive an earthquake, but will be damaged \cite{ref-12}.
Surprisingly, a study published by FERC indicated that the most damaging force to these dams are the local seismic activities and not a CSZ event \cite{ref-13}.
Although a dam structure may survive, significant damage inside can affect the generators, which would require a lengthy repair process.

\section{Transmission Level}

Transmission lines, substations, and substation equipment make-up a large fraction of electrical infrastructure. These critical pieces of equipment cross the PNW, covering great distances; further increasing the risk of damage from an earthquake.

Landslides pose the greatest risk to transmission lines and towers that cross mountain ranges: they are very likely to occur during an earthquake \cite{ref-08}, and given the length of some of the transmission lines the likelihood of lines being down is a significant factor. This is especially true for the connections crossing the Coastal Mountain range and Cascade Mountain range. 
An examination of the Oregon coast shows that very little generation lies on the coastal side of the mountains. This requires the power to be transferred across at least one mountain range, and increases the likelihood of critical line failures.

Another consideration of the transmission lines crossing mountain ranges, is that a large earthquake would render them nearly impossible to access by ground transportation.
The Oregon Resilience Plan highlighted this as a concern and the impacts were demonstrated in Nepal \cite{ref-01, ref-14}.
Minimal damage levels would damage bridges and shift or settle roads, making most routes impassible. In extreme cases, landslides may completely block routes. Nepal had several remote sites that were inaccessible following the earthquake, which made repairing some of its generation facilities difficult \cite{ref-14}.
Following a CSZ event, restoration efforts will be slowed due to blocked routes.

Soil liquefaction, settling, and lateral spreading also threaten transmission lines, towers, substations and substation equipment.
In the Christchurch, New Zealand earthquake, several sites showed signs of soil liquefaction and settling \cite{ref-11}. Some of these sites were abandoned in favor of rebuilding in new areas less susceptible to liquefaction. At one site, liquefaction tipped transformers at a noticeable angle, leading to a complete loss of the substation.
Many of New Zealand’s substations were originally unreinforced masonry, but a significant effort started years prior, reinforced the substations. This is considered a significant factor in almost all the substations surviving the earthquake \cite{ref-11}.

Although many models are available for predicting soil liquefaction, determining specific movements requires an in-depth site analysis.
In the PNW, a site analysis was performed at a substation near the coast. Two bore holes were drilled to determine the depth of the bedrock. The bedrock was at significantly different depths at opposite ends of the same substation. The analysis concluded that soil liquefaction is a primary concern at this site, and it would be a significant process to mitigate the risk.
This is likely the situation at most sites west of the Cascade Mountains, where ground shaking is likely to be high from a CSZ earthquake.

\section{Distribution Level}

Threats to the distribution system are the most well documented, and often the most heavily damaged in terms of downed pieces of equipment.
Few significant earthquakes have occurred in the PNW making a direct study difficult.
The Nisqually and New Zealand earthquakes, while lesser in magnitude, provide a detailed look at the threat of a smaller earthquake.

Soil liquefaction, settling and lateral spreading proved to be the most damaging to the distribution level equipment.
In many areas, buried cables are used instead of overhead power lines.
Early reports in New Zealand stated that 4\% of the 11kV buried cables failed, but this number was later increased to 15\% because of continued failures after being reenergized \cite{ref-11}.
The study of the New Zealand earthquake showed that buried cables were particularly susceptible to damage from liquefaction. In many of the instances in this earthquake, concrete failed putting additional strain on the buried cables. The strain caused high curvature bends and stretching. Later, when re-energized, the weakened cables failed anywhere from seconds to months following the earthquake \cite{ref-11}.
The buried cables also made identifying and repairing failures more difficult and time consuming, requiring a minimum of 12 hours to locate and repair each fault \cite{ref-11}.

In the Nisqually earthquake, it was observed that most of the outages occurred near the Bayfront, around Harbor Island and south to the Seattle International Airport \cite{ref-15}.
This area also experienced the greatest amount of ground shaking. The report compared the recorded shaking to the city projections, and showed issues created by artificial land and in saturated soils. Harbor Island is a man-made island built from the hills that used to lie near the Bayfront.
In Japan, reclaimed and coastal areas experienced the greatest amount of settling and soil liquefaction as well \cite{ref-10}. However, in areas of Japan where liquefaction prevention measures were in place, liquefaction was far less of an issue \cite{ref-10}.
In Seattle, half of the feeders in the severely shaken regions experienced failures. It was indicated that the failures were likely the result of base movement in poles near the artificially created areas of Harbor Island and the Duwamish River \cite{ref-15}. 37\% of this damage fit the previously mapped liquefaction zones for Seattle.

In larger earthquakes, like Chile, distribution level equipment had issues beyond soil and ground shaking \cite{ref-16}.
The main difficulties in restoring distribution level power was that several buildings fell on lines and the tsunami washed away lines and poles. Thus, power restoration in heavily damaged areas took weeks as new poles and lines were put in place \cite{ref-16}.
In Chile and New Zealand, restoration efforts were helped significantly by temporary generators \cite{ref-11, ref-16}.
New Zealand used the diesel generators to power critical loads, such as water facilities \cite{ref-11}.
Chile, used them in more isolated areas \cite{ref-16}.

In both buried and above ground distribution networks, soil liquefaction proved to be an issue.
Buried lines in liquefiable soils experienced concrete fracturing and ground displacement causing lines to fail.
Above ground lines suffered from having poles tip in poor soils or by being destroyed by falling structures.
The shaking itself did little harm to distribution level equipment.

\section{Conclusion}

In this paper, the current state of resilience in the state of Oregon was presented, along with systematized power system related experiences of historical earthquakes, proven practices and innovative ideas from affected regions, in order to contribute to the preparation for a future CSZ event in the PNW.

This paper is intended to encourage and catalyze the creation of a more resilient U.S.~PNW and Oregon electrical grid. Better resilience can be achieved, but the threat first must to be recognized at all levels and appropriate attention must be paid to the issue. Although the research is still in early phases, the move toward addressing this critical issue has fortunately already begun.

\vspace{0.1in}
\noindent
\textbf{Acknowledgements} This research was supported by the Oregon Talent Council under the \emph{``Pacific Northwest Electrical System Resiliency and Disaster Preparedness Training Project"}.
The authors would like to thank OTC for their financial support, and would also like to thank project and industry partners: Portland State University, Central Lincoln PUD, Portland General Electric, and Pacific Power.

\end{document}